# Substrate effect on the resistive switching in BiFeO$_3$ thin films

Yao Shuai,[1,2] Xin Ou,[1] Chuangui Wu,[2] Wanli Zhang,[2] Shengqiang Zhou,[1] Danilo Bürger,[1] Helfried Reuther,[1] Stefan Slesazeck,[3] Thomas Mikolajick,[3] Manfred Helm,[1] and Heidemarie Schmidt[1]

[1]*Institute of Ion Beam Physics and Materials Research, Helmholtz-Zentrum Dresden-Rossendorf, P. O. Box 510119, Dresden 01314, Germany*
[2] *State Key Laboratory of Electronic Thin Films and Integrated Devices, University of Electronic Science and Technology of China, Chengdu 610054, China*
[3]*Namlab gGmbH, Nöthnitzer Strasse 64, 01187 Dresden, Germany*

Abstract: BiFeO$_3$ thin films have been deposited on Pt/sapphire and Pt/Ti/SiO$_2$/Si substrates with pulsed laser deposition using the same growth conditions, respectively. Au was sputtered as the top electrode. The microscopic structure of the thin film varies by changing the underlying substrate. Thin films on Pt/sapphire are not resistively switchable due to the formation of Schottky contacts at both the top and the bottom interface. However, thin films on Pt/Ti/SiO$_2$/Si exhibit an obvious resistive switching behavior under forward bias. The conduction mechanisms in BiFeO$_3$ thin films on Pt/sapphire and Pt/Ti/SiO$_2$/Si substrates are discussed to understand the different resistive switching behaviors.

1. INTRODUCTION

Resistive switching devices have been intensively studied in recent years due to the advantages such as high switching speed, simple fabrication processes, and upscaling possibilities. Interesting resistive behavior has been observed in various binary and ternary oxides, for example, in $TiO_2$,[1] $NiO$,[2] $ZnO$,[3] $Pr_{0.7}Ca_{0.3}MnO_3$,[4] and $SiTiO_3$.[5] The mechanisms of the resistive switching are categorized into two types, i.e., filament[1,2] and interface switching.[4] However, the underlying physical origin of filament formation or interface switching is still under debate, and various models have been proposed to interpret the observed resistive switching in different materials. Generally, the ion migration is believed to cause filament formation or interface switching.[1,5] The movement of ions under an external voltage leads to the rupture and formation of local conductive filaments, or modifies the barrier height at the interface. Moreover, the electron trapping has also been reported to play an important role in resistive switching,[4] especially in the type of interface switching. The trapping or detrapping of electrons on those trapping sites changes the contact barrier and induces a homogeneous resistive switching at the interface.

We have previously reported on resistive switching in $BiFeO_3$ (BFO) thin films, and attributed it to an electron trapping effect.[6,7] In the present work, BFO thin films were deposited on Pt/sapphire and on $Pt/Ti/SiO_2/Si$ substrates. By comparing the conduction mechanisms of those BFO thin films on different substrates, the physical origin of the resistive switching in our BFO thin films is clarified.

2. EXPERIMENT

The BFO thin films were deposited with pulsed laser deposition at 600 °C, while keeping the oxygen pressure at 13 mTorr. The substrates were chosen to be $Pt/Ti/SiO_2/Si$ and Pt/sapphire.

Scanning Electron Microscope (SEM) was used to examine the microstructure of the BFO on different substrates. For electric measurements, Au top electrodes with an area of 0.1 mm$^2$ were sputtered using a metal shadow mask. The current-voltage (I-V) curves were recorded with a Keithley 2400 source meter.

3. RESULTS AND DISCUSSIONS

The surface morphology images measured by AFM on the samples after heating the substrate and with and without depositing the BFO thin film are shown in Fig. 1. The BFO thin film on Pt/sapphire exhibits an average grain size of ~120 nm [Fig. 1(a)], which is much smaller than that on Pt/Ti/SiO$_2$/Si [Fig. 1(c)]. Moreover, the BFO surface roughness is only 1 nm on Pt/sapphire, while it is increased to 12.2 nm by using Pt/Ti/SiO$_2$/Si substrates. The apparent change of the BFO topography is due to the distinct bottom electrode surface after the thin film growth at high temperature. As shown in Fig. 1(b) and (d), the surface roughness of the Pt bottom electrode amounts to 1.4 nm and 10.4 nm on sapphire and Ti/SiO$_2$/Si, respectively. The morphology of the Pt surface significantly influences the electrical properties of the BFO thin films, which will be discussed later.

The SEM cross section images are shown in Fig. 2. As expected for the same PLD growth conditions, the BFO thin films have a similar thickness of ~500 nm on both substrates. A sharp interface between the film and the Pt bottom electrode can be observed on sapphire substrates. However, the interface is rather rough for BFO on Pt/Ti/SiO$_2$/Si substrate, which is likely due to the interdiffusion caused by the high growth temperature. High quality interfaces on Pt/sapphire substrates have also been reported in chemical solution prepared BFO thin films.[8] It is noticed that the columnar grains extend from the bottom electrode throughout the whole thin film on Pt/sapphire [Fig. 2(a)], however, the film on Pt/Ti/SiO$_2$/Si consists of multi-layer grains [Fig. 2(b)].

I-V curves of the BFO thin films on Pt/sapphire and Pt/Ti/SiO$_2$/Si are shown in Fig. 3(a). Although the BFO growth conditions and the evaporation of the top contact are the same, significant differences can be observed in the I-V curves. First of all, the thin film on Pt/sapphire shows a symmetric I-V curve, while that on Pt/Ti/SiO$_2$/Si has a rectifying I-V characteristic. Moreover, the leakage current of the former one is much lower in positive bias range. In the negative bias, the leakage currents of the BFO films on different substrates are comparable. Finally and also most interesting, only the thin film on Pt/Ti/SiO$_2$/Si shows a hysteresis behavior, indicating that it is resistively switchable. The positive bias sets the film to LRS, and the negative bias resets it back to HRS.

To understand the difference of the I-V characteristics, it is necessary to investigate the conduction mechanisms of those thin films. As shown in Fig. 3(b), linear fittings are obtained for the film on Pt/sapphire substrates in both voltage polarities in a log(J) ~ E$^{0.5}$ scale. Note that only one branch is fitted for each voltage polarity, since there is no hysteresis. However, for the film on Pt/Ti/SiO$_2$/Si substrate, linear fitting is only possible in negative bias range [Fig. 3(b)]. By using the Schottky emission equation:

$$J = A^* T^2 \exp-(\frac{\varphi_b}{kT} - \frac{q}{kT}\sqrt{\frac{qE}{4\pi\varepsilon_0 K}}) \quad (1)$$

optical dielectric constants (K) are calculated to be 5.91 (positive bias) and 6.39 (negative bias) for the film on Pt/sapphire, and 6.85 (negative bias) on Pt/Ti/SiO$_2$/Si. These values are in good agreement with the previous report that K is 6.25 for BFO thin films,[9] indicating that the Schottky emission is the dominant conduction mechanism in these voltage ranges for different samples.

Therefore, the symmetric I-V curve of the thin film on Pt/sapphire substrate is likely due to the formation of Schottky contacts at both interfaces,[10] while the rectifying behavior of the thin film on Pt/Ti/SiO2/Si is because of the asymmetric contact barriers at the two interfaces. Note that the comparable leakage current in the negative bias range for the thin films on different substrates is likely due to the same conduction mechanism when the interface between BFO thin film and top contact is involved, i.e. the Schottky junction at the top interface shows a very similar conduction behavior.

However, an obvious hysteresis behavior is observed in the positive bias range of the thin film on Pt/Ti/SiO$_2$/Si substrate [Fig. 3(a)]. In contrast to the Schottky emission in the negative bias range [Fig. 3(b)], two different types of conduction mechanisms dominate the upper (LRS) and lower (HRS) I-V branches, respectively. The lower branch, i.e. the HRS, can be well fitted using the Poole-Frenkel model, which deduces an optical dielectric constant (K) of 5.02 [Fig. 3(c)]. Since PF conduction is bulk-limited, the contact resistance of the bottom interface is much smaller than the bulk thin film, and the bottom contact behaves like an Ohmic contact, which results from the rough bottom Pt surface [Fig. 1(d)] and the interdiffusion at the BFO/Pt interface [Fig. 2(b)] Note that although the top interface is a Schottky contact, it is not possible to obtain a reasonable fitting by using the forward Schottky emission model. Therefore, the resistance of the top interface is also negligible as compared to the bulk thin film when it is forward biased. However, when the voltage is large enough, the fitting deviates from the PF model, and tunneling starts to appear as indicated by the negative slope of the linear fitting in a $\log(J/E^2) \sim 1/E$ scale [Fig. 3(d)]. Furthermore, the upper branch (LRS) is also dominated by tunneling as revealed by the fitting [Fig. 3(d)]. This suggests that when the thin film is set to LRS, the bulk resistance of the thin film is significantly decreased, and the interface resistance (tunneling effect) becomes dominant. Because the bottom interface is Ohmic contact, tunneling takes place at the top interface, revealing that the depletion region at the Au/BFO interface is considerably narrowed. Note that the small hysteresis in the negative bias range is likely due to the thin depletion thickness induced by the positive bias, the deviation of the fitting in Fig. 3(b) can be attributed to the presence of tunneling through the depletion layer. By further increasing the negative bias, the depletion region is extended, which eliminates the tunneling, and only Schottky emission dominates.

As discussed above, the resistive switching of the thin film on Pt/Ti/SiO$_2$/Si substrate is due to the nonvolatile modification of the depletion thickness at the top interface (Au/BFO) together with the resistance change of the bulk thin film. However, the physical origin of this nonvolatile effect needs to be clarified. Ion migration model has been proposed in a wide range of resistive switching behaviors in various materials. However, when taking into account the polarities for the two resistance states, the ion migration model is not applicable. The most movable ions in BFO thin films are oxygen vacancies (OVs), which are donors since they provide electrons.[11] If a positive bias is applied on the top Au electrode, the OVs are pushed away from the Au/BFO interface since they are positively charged. That should raise the barrier height, because the concentration of donors is decreased at the interface. Thus the positive bias sets the structure to HRS if the ion migration model dominates, which is just the opposite of our observation as shown in Fig. 3(a). By comparing the I-V characteristics of those thin films on different substrates, the nonvolatile resistive switching is likely due to a pure electronic effect, i. e. the electron trapping effect. No resistive switching has been observed in the thin film on Pt/sapphire substrate, because the two Schottky interfaces block the electron injection in both voltage polarities. And only the thin film on Pt/Ti/SiO$_2$/Si shows an obvious resistive switching in the positive bias range, due to the large amount of injected electrons through the Ohmic BFO/Pt contact.

4. CONCLUSIONS

In summary, BiFeO$_3$ thin films have been deposited on Pt/sapphire and Pt/Ti/SiO$_2$/Si substrates, respectively. The Schottky contacts of the thin film on Pt/sapphire dominate the transport and help to suppress the leakage current in both voltage polarities, and the I-V curve shows no hysteresis due to the reversely biased Schottky contact. However, the thin film on Pt/Ti/SiO$_2$/Si shows a diode behavior and is resistively switchable, which results from the asymmetric contact geometry at the two interfaces. By comparing the electric properties of the thin films on different substrates, the mechanism of the observed resistive switching is clarified to be due to an electron trapping effect.

ACKNOWLEDGMENTS

Y. S. would like to thank the China Scholarship Council (grant number: 2009607011). S. Z. acknowledges the funding by the Helmholtz-Gemeinschaft Deutscher Forschungszentren (HGF-VH-NG-713). D. B. and H. S. thank the financial support from the Bundesministerium für Bildung und Forschung (BMBF grant number: 13N10144).

Figure captions:

Fig. 1. AFM surface images of the heated Pt/sapphire and Pt/Ti/SiO$_2$/Si substrates with (a,c) and without (b,d) BFO thin film growth.

Fig. 2. SEM cross section images of the BFO thin films on Pt/sapphire (a), and on Pt/Ti/SiO$_2$/Si (b).

Fig. 3. (a) I-V curves of the BFO thin films on Pt/sapphire and on Pt/Ti/SiO2/Si substrates, respectively. (b) Log J ~ E$^{0.5}$ scale representing Schottky emission. (c) Log (J/E) ~ E$^{0.5}$ scale representing Poole-Frankel conduction. (d) Log (J/E$^2$) ~ 1/E scale representing tunneling.

Figures

Fig. 1

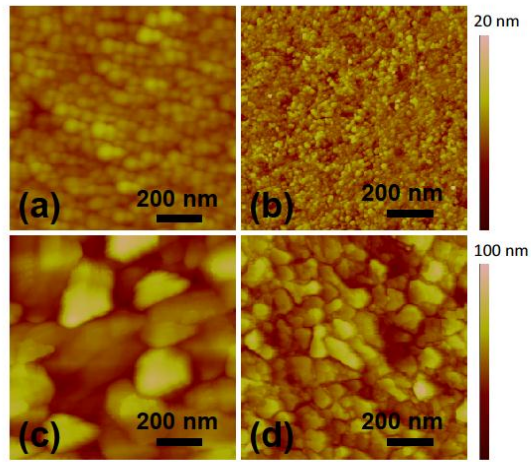

Fig. 2

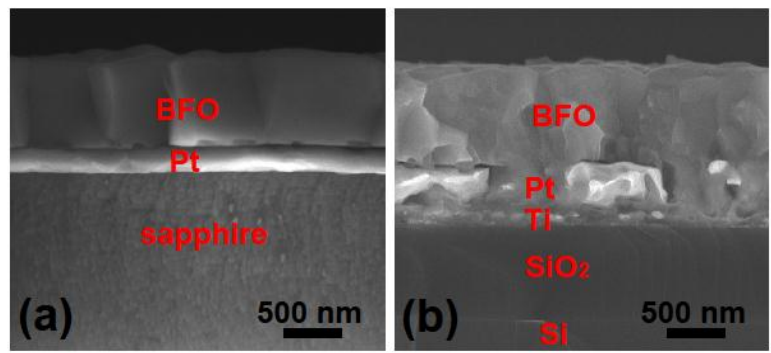

Fig. 3

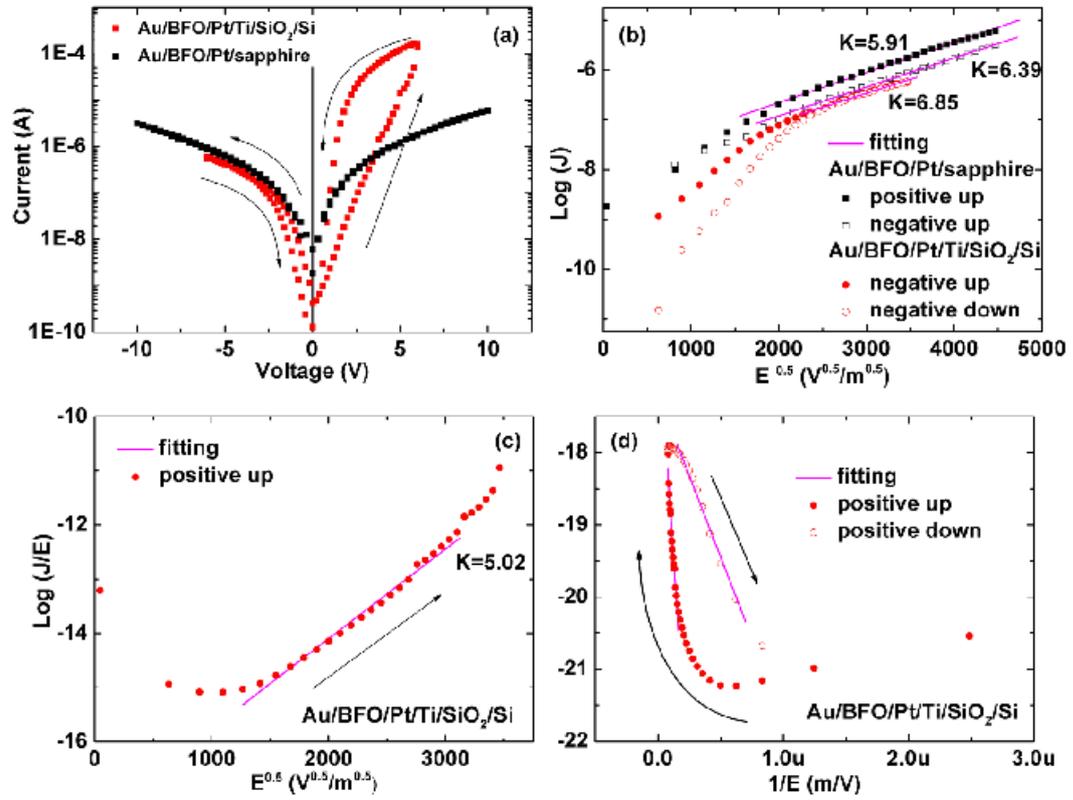